\documentclass[aps,prl,preprint,groupedaddress]{revtex4-1}

\usepackage{amsmath}
\usepackage{amssymb}
\usepackage{graphicx}
\usepackage{mathrsfs}
\usepackage{lineno}
\usepackage{bbm}
\usepackage{mathbbol}
\usepackage{tabularx} 
\usepackage{multirow} 
 
\begin{document}
\raggedbottom

\title{A tau-leaping method for computing joint probability distributions of the first-passage time and position of a Brownian particle.}

\author{Jaroslav Albert}

\begin{abstract}
First passage time (FPT), also known as first hitting time, is the time a particle, subject to some stochastic process, 
hits or crosses a closed surface for the very first time. $\tau$-leaping methods are a class of stochastic algorithms in which,
instead of simulating every single reaction, many reactions are ``leaped" over in order to shorten the computing time. 
In this paper we developed a $\tau$-leaping method for computing the FPT and position 
in arbitrary volumes for a Brownian particle governed by the Langevin equation. The 
$\tau$-leaping method proposed here works as follows. A sphere is inscribed within the volume of interest (VOI) centered at the initial particle's location. On this sphere, the FPT is sampled,
as well as the position, which becomes the new initial position. Then, another sphere, centered at this new location, is inscribed. This process continues until the 
sphere becomes smaller than some minimal radius $R_{\text{min}}$. When this occurs, the $\tau$-leaping switches to the conventional Monte Carlo, which runs until the particle either crosses the surface
of the VOI or finds its way to a position where a sphere of radius $>R_{\text{min}}$ can be inscribed. The switching between $\tau$-leaping and MC continues until the particle
crosses the surface of the VOI. The purpose of a minimal radius is to avoid having to sample the velocities, which become irrelevant when the particle diffuses beyond a certain 
distance, i. e. $R_{\text{min}}$ The size of this radius depends on the system parameters and on one's notion of accuracy: the larger this radius the more accurate the $\tau$-leaping method, but also
less efficient. This trade off between accuracy and efficiency is discussed. For two VOI, the $\tau$-leaping method is shown to be 
accurate and more efficient than MC by at least a factor of 10 and up to a factor of about 110. However, while MC becomes exponentially slower with increasing VOI, 
the efficiency of the $\tau$-leaping method remains relatively unchanged. Thus, the $\tau$-leaping method can potentially be many orders of magnitude more efficient than MC. 
\end{abstract}

\maketitle

\section{Introduction}

First passage time (FPT) is the time that a certain event occurs for the first time during an evolution of a system.
In molecular biology it is often desirable to know the FPT distributions for a molecule, such as protein, for crossing a surface, e. g. that of the cell nucleus or the cell membrane, or
for finding its target site on the DNA \cite{Chou}.
Although mean first passage times for these types of events have been worked out to various degrees of approximations \cite{Singer, Ward}, the non-trivial shapes of volumes and obstacle-riddled environments
in which biological molecules have to navigate makes computations of FPT distributions difficult. The usual strategy in such efforts is to simulate the molecular dynamics using Monte Carlo methos, which
do get the job done but are notoriously inefficient.

In this paper we draw inspiration from computational analysis of stochastic gene expression -- an area of research that has produced many alternative methods to brute Monte Carlo simulations.
In particular, we focus on two such methods: $\tau$-leaping \cite{Gillespie, Rathinam, Tian, Chatterjee, Cao, Pettigrew, Anderson, Hu, Anderson2, Lago, Rossinelli, Koh, Padgett} and hybrid stochastic simulation algorithms (HSSA) \cite{Haseltine, Rao, Burrage, Salis, Weinan, Samant, Salis2, Jahnke, Zechner, Albert, Albert2, Duso, Kurasov, Albert3}. A $\tau$-leaping method approximates the evolution of a system over many small steps
in a MC simulation by taking larger steps or leaps, thereby reducing the overall number of steps that need to be taken. The HSSAs on the other hand, work by employing a form of $\tau$-leaping method on a part of the system
(a subset of molecular species and chemical reactions), while using good old MC on the rest of the system. In this paper we apply these concepts to Brownian motion described by the Langevin equation
in volumes of arbitrary shapes with the goal to compute joint distributions of the FPT and position.
More specifically, we take advantage of the fact that LEs can be solved approximately for a spherical volume of certain minimal size, which can be used to fill parts of the larger volume of interest. Sampling
the FPT and position for this spherical volume, we generate another sphere centered at the sampled position. When this process
brings the particle within a certain distance from the boundary, we switch to the MC. Thus, with each sphere, we effectively $\tau$-leap over $\tau/dt$ number of steps, where $dt$ is the temporal size of each step in the MC simulation.
With full details about what happens near the boundary, we show the accuracy and efficiency of our method on two examples volumes.

\section{Brownian motion and the Langevin equation} 

When a large particle is immersed in a medium (gas or liquid) of many smaller particles at equilibrium, it moves in a jittery fashion due to density fluctuations in that medium. One
model of such motion is called Brownian, and is described by the Langevin equation,
\begin{eqnarray}
&&m\frac{d{\bf v}(t)}{dt}-\frac{{\bf v}(t)}{\tau_B}={\bf f}(t)\label{Langevin}
\end{eqnarray}
where ${\bf v}(t)$ is the particle's velocity at time $t$, $m$ is its mass, $\tau_B$ is the relaxation time, and
${\bf f}(t)$ is a random force. This random force changes magnitude and direction at time intervals separated by $dt$ and follows a Gaussian distribution
\begin{equation}\label{fdistribution}
P({\bf f})=\frac{1}{(2\pi\sigma_f^2)^{3/2}}e^{-{\bf f}\cdot{\bf f}/(2\sigma_f^2)},
\end{equation}
where $\sigma_f^2=2k_BTm/(dt\tau_B)$, and $k_B$ and $T$ are the Boltzman constant and temperature, respectively. The relaxation time $\tau_B$ is related to the mass $m$,
viscosity of the medium $\nu$, and the particle's size $r_B$ via this expression: $\tau_B=m/(6\pi\nu r_B)$.
Hence, coupled with the definition of velocity, ${\bf v}=d{\bf r}/dt$, Eq. (\ref{Langevin}) can be used to simulate the evolution of a Brownian particle's velocity and position ${\bf r}$ by
iteration. The time step, $dt$, must be chosen to satisfy $\tau_s\ll dt$, where $\tau_s$ is the average collision time between the Brownian particle and the molecules of the medium.

Another approach to studying Brownian motion is via a Master Equation for
the joint probability distribution, $P({\bf r},{\bf v},t)$, which is given by the Klein-Kramers equation (also referred to as Fokker-Planck equation) \cite{Kramers}: 
\begin{equation}\label{Kramers}
\frac{\partial P}{\partial t}+{\bf v}\cdot{\bf \nabla}_{\bf r}P-\frac{1}{\tau_B}{\bf v}\cdot{\bf \nabla}_{\bf v}P-\frac{k_BT}{\tau_B m}{\bf \nabla}^2_{\bf v}P=0.
\end{equation}
The solution to Eq. (\ref{Kramers}) with infinite boundaries and the initial conditions $P({\bf r},{\bf v},0)=\delta^{(3)}({\bf r}-{\bf r'})\delta^{(3)}({\bf v}-{\bf v'})$ is given by \cite{Chandrasekhar, Risken}:
\begin{eqnarray}\label{KKsolution}
P({\bf r},{\bf v},t)&=&\frac{1}{(2\pi\sigma_X\sigma_V\sqrt{1-\beta^2})^3}\times\nonumber\\
&&\text{exp}\left[-\frac{1}{2(1-\beta^2)}\left(\frac{|{\bf r}-{\pmb\mu}_X|^2}{\sigma_X^2}+\frac{|{\bf v}-{\pmb\mu}_V|^2}{\sigma_V^2}
-\frac{2\beta({\bf r}-{\pmb\mu}_X)\cdot({\bf v}-{\pmb\mu}_V)}{\sigma_X\sigma_V}\right)\right],
\end{eqnarray}
where
\begin{eqnarray}
&&\sigma_X^2=\frac{k_BT\tau_B^2}{m}\left[1+2t/\tau_B-\left(2-e^{-t/\tau_B}\right)^2\right]\label{sigX}\\
&&\sigma_V^2=\frac{k_BT}{m}\left(1-e^{-2t/\tau_B}\right)\label{sigV}\\
&&\beta=\frac{k_BT\tau_B}{\sigma_X\sigma_V}\left(1-e^{-t/\tau_B}\right)^2\nonumber\\
&&{\pmb\mu}_X={\bf r'}+(1-e^{-t/\tau_B})\tau_B{\bf v'}\nonumber\\
&&{\pmb\mu}_V={\bf v'}e^{-t/\tau_B}\nonumber.
\end{eqnarray}

We can obtain the probability for the particle's position by integrating Eq. (\ref{KKsolution}) over ${\bf v}$:
\begin{equation}\label{P(x)}
P({\bf r},t)=\int_{-\infty}^\infty P({\bf r},{\bf v},t)d{\bf v}=\frac{1}{(2\pi\sigma_X(t)^2)^3}\text{exp}\left[-\frac{|{\bf r}-{\pmb\mu}_X(t)|^2}{2\sigma_X(t)^2}\right].
\end{equation}
For $t\gg\tau_B$, $\sigma_X(t)^2\rightarrow2(k_BT/m)t$ and ${\pmb\mu}_X(t)\rightarrow{\bf r'}+\tau_B{\bf v'}$, which allows us to replace the Brownian model with a diffusion model:
\begin{equation}\label{diff-model}
\frac{\partial P({\bf r},t)}{\partial t}=\frac{1}{D}\nabla^2P({\bf r},t),
\end{equation}
where $D=k_BT/m$, subject to the initial conditions $P({\bf r},0)=\delta^{(3)}({\bf r}-{\bf r'}-\tau_B{\bf v'})$. 
We can quantify the discrepancy between the Langevin model
and the diffusion model via this expression:
\begin{equation}\label{d}
w(t)=1-\frac{\sigma_X(t)^2}{2Dt}.
\end{equation}
If we set $w(t)$ to some small value $\varepsilon_w$, we can solve Eq. (\ref{d}) for the minimal time the system must evolve before we can treated as diffusive: $t_{\text{min}}=3\tau_B/(2\varepsilon_w)$. For example, if $\varepsilon_w=0.03$,
we get $t_{\text{min}}=50\tau_B$. Thus, if we are only interested in times $>t_{\text{min}}$, we are free to use Eq. (\ref{diff-model}) as our model. Although one can chose ${\bf v'}$ in the initial conditions to be any value, it is useful to consider the magnitude of the term $\tau_B{\bf v'}$ for a realistic scenario, e. g.
${\bf v'}$ being the result of a Brownian particle having arrived at position ${\bf r'}$ at time $t=0$, after traveling for a time $>t_{\text{min}}$. According to Eq. (\ref{sigV}), the distribution of velocities for such a particle would have
the standard deviation $\sigma_V^2=k_BT/m$. Thus, the maximum speed of the arriving Brownian particle would be $\sim3\sqrt{k_BT/m}$. For a large enough volume, we can assume the term $3\tau_B\sqrt{k_BT/m}$ to be negligible, i. e. if
$3\tau_B\sqrt{k_BT/m}/R\ll1$, where $R$ is the radius of our sphere. By choosing the smallness of $\varepsilon_R=3\tau_B\sqrt{k_BT/m}/R$, e. g. $\varepsilon_R=0.03$, we can determine the minimum radius $R$ for which the initial velocity can be neglected:
$R_{\text{min}}=3\varepsilon_R\sqrt{m/(k_BT)}\tau_B$. Thus, provided the particle takes significantly longer on average than $t_{\text{min}}$ to reach a distance $R_{\text{min}}$,
we can replace the Brownian model with a diffusion model for $t>t_{\text{min}}$. In a moment we will see that the minimal time to reach a distance $R_{\text{min}}$ is indeed much longer than $t_{\text{min}}$.
With these criteria we can compute the FPT for a Brownian particle using Eq. (\ref{diff-model}) and the initial condition $P({\bf r},0)=\delta^{(3)}({\bf r})$.
In spherical coordinates, Eq. (\ref{diff-model}) reads:
\begin{equation}
\frac{\partial P(\xi,t)}{\partial T}=\frac{1}{\xi^2}\frac{\partial}{\partial\xi}\left(\xi^2\frac{\partial P(\xi,t)}{\partial\xi}\right),
\end{equation}
where $\xi=r/R$ and $T=Dt/R^2$. Thanks to spherical symmetry, $P({\bf r},t)$ is independent of the longitudinal and azimuthal angles, $\phi$ and $\theta$. 
The initial condition becomes $P({\bf r},0)=\delta(r)/r^2$, or $P(\xi,0)=\delta(\xi)/\xi^2$.
To compute the FPT, we also need to add the absorbing boundary condition $P(\xi=1,T)=0$, for which the solution is:
\begin{equation}
P(\xi,T)=\lim_{M \to\infty}P_M(\xi,T),
\end{equation}
where\begin{equation}
P_M(\xi,T)=\sum_{n=1}^M A_n(M)\frac{\sin(\pi n\xi)}{\xi}e^{-(\pi n)^2T}
\end{equation}
and
\begin{equation}
A_n(M)=2\pi ne^{-(\pi n/M)^2}.
\end{equation}
The survivor's probability $S_{\infty}(T)$, which is the probability that the particle remains inside $R$ for a period of time $T$, is given by
\begin{equation}
S_{\infty}(T)=\int_0^1 P_M(\xi,T)\xi^2d\xi=\lim_{M \to\infty}\sum_{n=1}^M2 (-1)^{n+1}e^{-(\pi n)^2(T+1/M)}.
\end{equation}
The subscript ${\infty}$ serves as a reminder that $M\rightarrow\infty$. 
The FPT distribution is simply $F_\infty(T)=1-S_\infty(T)$.
In practice, however, the summation limit can be cut off at some finite value of $M$: $F_M(T)=1-S_M(T)$. 
Since the exponential term $e^{-(\pi n)^2(T+1/M)}$ decays very rapidly for large $n$, we can take the limit $(T+1/M)\rightarrow T$, while
cutting the summation off at some finite $M$ to obtain:
\begin{equation}
{\tilde F}_M(T)=1-\sum_{n=1}^M2 (-1)^{n+1}e^{-(\pi n)^2T}.
\end{equation}
Figures 1 a) and b) show the behaviors of $P_M(\xi,0)$, $F_M(T)$ and${\tilde F}_M(T)$  for different values of $M$.
Evidently, to make $P_M(\xi,0)$ sharply peaked near $r=0$ and $F_M(T)$ converge to $F_{\infty}(T)$ requires large values of $M$, while the function ${\tilde F}_M(T)$ does not: for $M=10$ it already behaves correctly
for $T$ larger than $\sim 0.004$.
\begin{figure}
	\centering
	\includegraphics[trim=0 0 0 1.0cm, height=0.175\textheight]{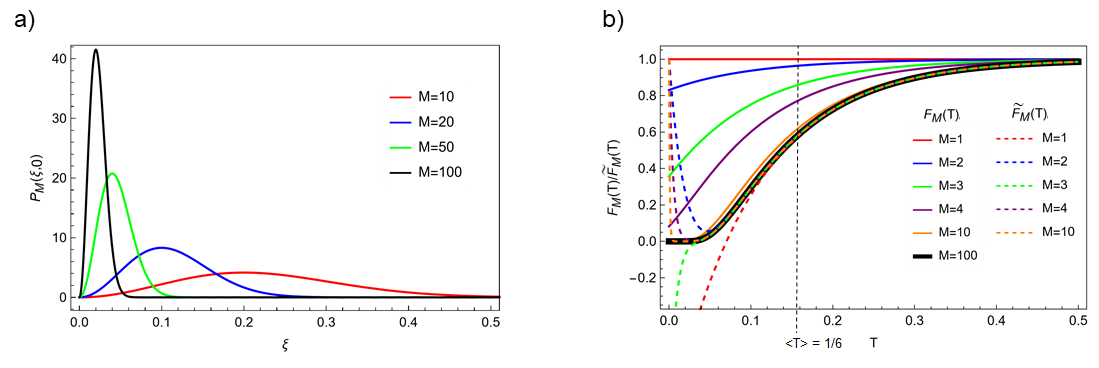}
	\caption{a) Initial distributions $P_M(\xi,0)$ for different $M$; b) Distributions $F_M(T)$ (solid lines) and ${\tilde F}_M(T)$ (dashed lines) for different $M$.}
\end{figure}

To sample ${\tilde F}_M(T)$, one needs only to generate a random real number $\eta=(0,1]$ and solve
${\tilde F}_M(T)-\eta=0$ for $T$. For nontrivial ${\tilde F}_M(T)$, this could be done by minimizing $|F_M(T)-\eta|$. However, if we compute the average $T$,
\begin{equation}
\langle T\rangle=\int_0^{\infty}\sum_{n=1}^{\infty}2 (-1)^{n+1}e^{-(\pi n)^2T}dT=\sum_{n=1}^{\infty}2\frac{(-1)^{n+1}}{(\pi n)^2}=\frac{1}{6},
\end{equation}
we notice, by examining Figure 1 b) again, that the function ${\tilde F}_1(T)=1-2e^{-\pi^2T}$ behaves correctly for $T>1/6$ - the average. Hence, we could speed up the
minimization procedure by first checking whether $1-2e^{-\pi^2/6}$ is greater or smaller than $\eta$. If it is the latter, we can set
$1-2e^{-\pi^2T}$ to $\eta$ and solve for $T$ analytically:
\begin{equation}
T=\frac{1}{\pi^2}\ln\left(\frac{2}{1-\eta}\right).
\end{equation}
If it is the former, we only need to search for $T$ in the range $[0,1/6]$.

Before we continue we must circle back and check that the time to reach a distance $R_{\text{min}}$ is much greater than $t_{\text{min}}$. We can do this by requiring that
${\tilde F}_{10}(T)$ be less than some chosen value, e. g. $0.001$, which corresponds to $T=0.04$. If we recall that $T=tMR_{\text{min}}/(k_BT\tau_B)$, $R_{\text{min}}= 3\tau_B\sqrt{k_B T/m}/\varepsilon_R$
and $t_{\text{min}}=3\tau_B/(2\varepsilon_w)$, we obtain $t/t_{\text{min}}=6\times0.04(\varepsilon_w/\varepsilon_R^2)$. For $\varepsilon_d=\varepsilon_R=0.03$, we get $t/t_{\text{min}}=8$.

\section{$\tau$-leaping}

\begin{figure}
	\centering
	\includegraphics[trim=0 0 0 1.0cm, height=0.4\textheight]{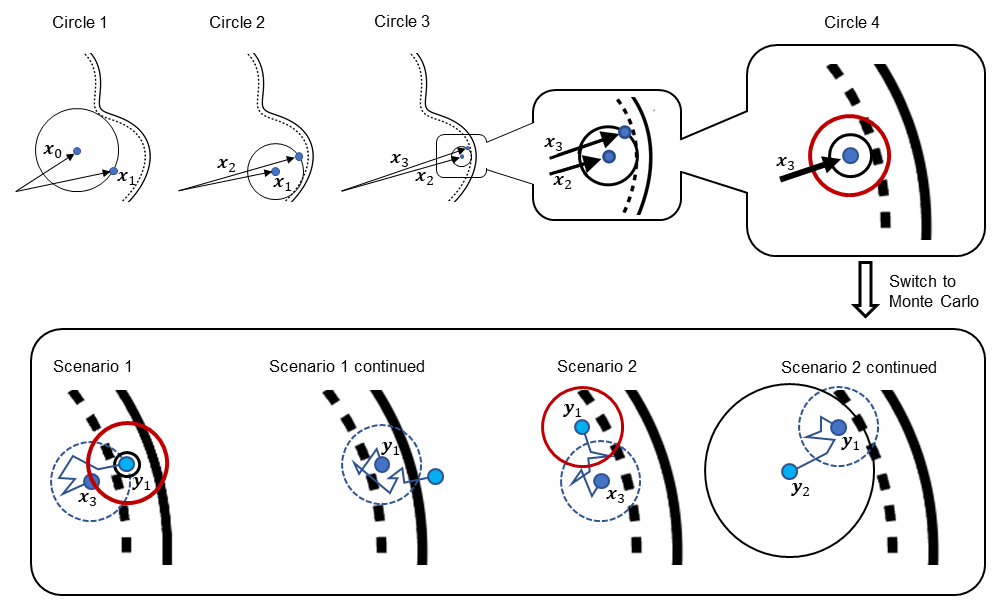}
	\caption{A 2-dimensional illustration of the $\tau$-leaping method. The particle starts out at position ${\bf x}_0$. The smallest possible circle centered at ${\bf x}_0$, 
Circle 1, is generated and a point on its surface, ${\bf x}_1$, is sampled. The same process is repeated for Circle 2, Circle 3 and Circle 4. However, the
radius of Circle 4 is smaller than $R_{\text{min}}$, so we must switch to Monte Carlo. In Scenario 1, the particle diffuses a distance $\ge R_{\text{min}}$ (blue dashed circle) to
the point ${\bf y}_1$;
a sphere centered at ${\bf y}_1$ is generated but its radius is smaller than $R_{\text{min}}$, so Monte Carlo continues until the particle crosses the boundary (black thick line). 
In Scenario 2, the particle diffuses a distance $\ge R_{\text{min}}$ to
the point ${\bf y}_1$; a sphere centered at ${\bf y}_1$ is generated with radius $<R_{\text{min}}$, so we continue with Monte Carlo; the particle
diffuses to a point ${\bf y}_2$, where a sphere of radius $>R_{\text{min}}$ is generated, and we switch to $\tau$-leaping to continue the process.}
\end{figure}


Now that we have an analytical expression for the FPT for a sphere, we can use it to speed up simulation of Brownian motion. The scheme is shown in Figure 2 on a two-dimensional example.
First, we give the surface of the volume of interest a skin of thickness $R_{\text{min}}$ on the inside (the purpose of which will be explained shortly). Next, starting from some initial point ${\bf x}_0$,
we generate a sphere centered at ${\bf x}_0$ such that its surface and the skin share a unique point. This is equivalent to finding the smallest sphere whose surface touches the skin. 
Then, we sample the FPT and the particle's position on the surface of the sphere, $(t_1,{\bf x}_1)$. 
Centered at ${\bf x}_1$, we generate another sphere whose surface touches the skin. We sample the FPT and surface position, $(t_2,{\bf x}_2)$, and continue this process in this manner
until we generate a sphere with a radius $>R_{\text{min}}$. When this happens, we switch to Monte Carlo, with the initial conditions ${\bf \tilde x}_0={\bf x}_i$ and ${\bf v}=0$. Of course, 
in reality there is no reason to expect ${\bf v}$ to be $0$,
unless we get very lucky. However, if we let the particle evolve past a radius $R_{\text{min}}$, we do not need to worry about its initial velocity and may set to zero. This is where the skin guarantees accuracy: in the
(unlikely) event of sampling a position that falls on the skin, we are guaranteed that, should the particle evolve past the outer surface, it will have traveled at least the distance $R_{\text{min}}$.
With this quality check in place, we can write down the steps of this procedure in more detail.

\vspace{-10mm}

\begin{eqnarray}
&&0:\,\,\,\,\,\text{Choose a volume whose enclosing surface is given by a vector ${\bf g}(\lambda_1,\lambda_2)$}, \nonumber\\
&&\,\,\,\,\,\,\,\,\,\,\,\,\,\text{parametrized by $\lambda_1$ and $\lambda_2$. Also choose $(T, m, \tau_B, R_{\text{min}})$ and the step size $dt$.}\nonumber\\
&&1:\,\,\,\,\,\text{Set $(p,n)=0$, where $p$ and $n$ are counters, and choose initial time $t_p$ (e. g. zero) and }\nonumber\\
&&\,\,\,\,\,\,\,\,\,\,\,\,\,\text{an initial position ${\bf x}_p$.}\nonumber\\
&&2:\,\,\,\,\,\text{Generate a sphere of radius $R$ by minimizing $|{\bf g}(\lambda_1,\lambda_2)-{\bf x}_p|$. If $R\ge R_{\text{min}}$, set} \nonumber\\
&&\,\,\,\,\,\,\,\,\,\,\,\,\,\text{$p=p+1$ and  go to step 3; otherwise go to step 6.}\nonumber\\
&&3:\,\,\,\,\,\text{Sample $T$ by generating a random real number $\eta=[0,1)$. If $\eta>1-2e^{\pi^2/6}$,}\nonumber\\
&&\,\,\,\,\,\,\,\,\,\,\,\,\,\text{set $T=1/\pi^2\ln[2/(1-\eta)]$; otherwise set $T=$min$|{\tilde F}_{10}(T')-\eta|$. Set $t_p=R^2T/D$ and}\nonumber\\
&&\,\,\,\,\,\,\,\,\,\,\,\,\,\text{record it.}\nonumber\\
&&4:\,\,\,\,\,\text{Sample a point on a sphere, ${\bf r}$, from a uniform distribution by generating two }\nonumber\\
&&\,\,\,\,\,\,\,\,\,\,\,\,\,\text{random numbers $q_1=[0,2\pi]$ and $q_2=[0,1]$ and set}\nonumber\\
&&\,\,\,\,\,\,\,\,\,\,\,\,\,\text{${\bf r}=[R\sin\theta\cos\phi, R\sin\theta\sin\phi, R\cos\theta]$, where $\theta=q_1$ and $\phi=\arccos(1-2q_2)$ \cite{Simon}.}\nonumber\\
&&5:\,\,\,\,\,\text{Set ${\bf x}_p={\bf x}_{p-1}+{\bf r}$ and go to step 2.}\nonumber\\
&&6:\,\,\,\,\,\text{Simulate Eq. (\ref{Langevin}) using Monte Carlo with the initial conditions ${\bf X}_n={\bf x}_p$ and ${\bf V}_n=0$,}\nonumber\\
&&\,\,\,\,\,\,\,\,\,\,\,\,\,\text{until a) ${\bf X}_n$ reaches the outside of the volume; or b) $|{\bf X}_n-{\bf x}_p|\ge R_{\text{min}}$.}\nonumber\\
&&\,\,\,\,\,\,\,\,\,\,\,\,\,\text{If a)  is satisfied, go to step 8; otherwise, set ${\bf x}_p={\bf X}_n$ and go to step 2.}\nonumber\\
&&8:\,\,\,\,\,\text{Record  ${\bf X}_n$ and the FPT $t=\sum_{i=0}^pt_i+ndt$.}\nonumber
\end{eqnarray}

\section{Validation}

In this section we apply our method to two example volumes and compare the results to Monte Carlo simulations. 

\subsection{Example 1}

\begin{figure}
	\centering
	\includegraphics[trim=0 0 0 1.0cm, height=0.32\textheight]{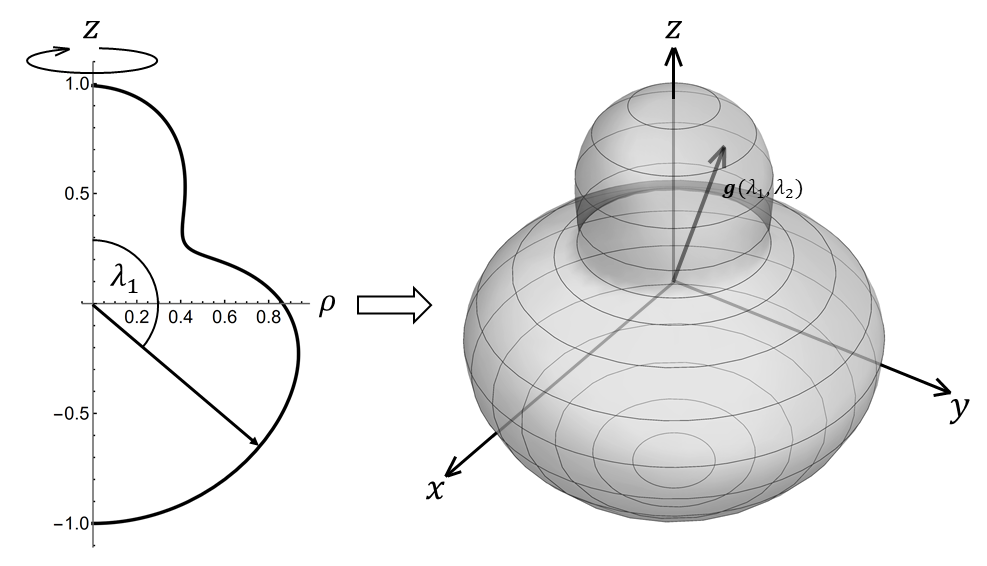}
	\caption{Volume of interest (right) generated by rotating a curve (left) around the z-axes. The parameters $\lambda_1$ and $\lambda_2$ play the role of 
the polar and azimuthal angle.}
\end{figure}

We chose a volume by revolving the curve 
\begin{equation}
h(\lambda_1)= 1-\frac{e^{-4 (\lambda_1-1)^2}}{2}.
\end{equation}
around the z-axes, where $\lambda_1$ has a range $[0,\pi]$. The corresponding volume is given by the vector 
\begin{equation}
{\bf g}(\lambda_1,\lambda_2)=(h(\lambda_1)\sin\lambda_1\cos\lambda_2,h(\lambda_1)\sin\lambda_1\sin\lambda_2,h(\lambda_1)\cos\lambda_1),
\end{equation}
shown in Fig. 3.

To give this volume a skin, we need to subtract $R_{\text{min}}{\bf u}(\lambda_1,\lambda_2)$ from ${\bf g}$, where ${\bf u}(\lambda_1,\lambda_2)$ is the 
unit vector perpendicular to the surface at the point $(\lambda_1,\lambda_2)$. Since the horizontal cross section of the volume is a circle, we can write ${\bf u}(\lambda_1,\lambda_2)$
and ${\bf g}(\lambda_1,\lambda_2)$ in cylindrical coordinates $(\rho,z,\phi)$, where  $\rho=\sqrt{x^2+y^2}$, as 
${\bf u}(\lambda_1)=(u_{\rho}(\lambda_1), u_z(\lambda_1),0)$ and ${\bf g}(\lambda_1,\lambda_2)=(f(\lambda_1)\sin\lambda_1,f(\lambda_1)\cos\lambda_1,g_{\phi}(\lambda_1,\lambda_2))$,
respectively. Finding $(u_{\rho}(\lambda_1)$ and  $u_z(\lambda_1))$ is then a matter of solving
the equation
\begin{equation}
\frac{d{\bf g}(\lambda_1,\lambda_2)}{d\lambda_1}\cdot{\bf u}(\lambda_1)=0,
\end{equation}
or
\begin{equation}
u_{\rho}[h(\lambda_1)\cos\lambda_1+h'(\lambda_1)\sin\lambda_1]+u_z(\lambda_1) [-h(\lambda_1)\sin\lambda_1+h'(\lambda_1)\cos\lambda_1]=0.
\end{equation}
Coupled with the condition that ${\bf u}(\lambda_1)$ has a unit length, i. e. $u_{\rho}^2+u_z^2=1$,
we obtain
\begin{eqnarray}
&&u_{\rho}(\lambda_1)= \frac{1}{\sqrt{1 + H(\lambda_1)^2}}\nonumber\\
&&u_z(\lambda_1)=-\frac{H(\lambda_1)}{\sqrt{1 +H(\lambda_1)^2}},\nonumber
\end{eqnarray}
where
\begin{equation}
H(\lambda_1)= \frac{h(\lambda_1)\cos\lambda_1+h'(\lambda_1)\sin\lambda_1}{ -h(\lambda_1)\sin\lambda_1+h'(\lambda_1)\cos\lambda_1}.\nonumber
\end{equation}
To generate a sphere centered at ${\bf x}_0$ that touches the skin at a single point, we only need to minimize its radius, or, equivalently, its square radius:
\begin{equation}
R(\lambda_1)^2=[h(\lambda_1)\sin\lambda_1-R_{\text{min}}u_{\rho}(\lambda_1)]^2+[h(\lambda_1)\cos\lambda_1-R_{\text{min}}u_z(\lambda_1)]^2.
\end{equation}

To sample $T$, we set it to
$(1/\pi^2)\ln(2/1-\eta)$ if $\eta>1-2e^{-\pi^2/6}$, otherwise we used the minimizer ``fminbnd" for the function $[{\tilde F}_{10}(T')-\eta]^2$ in the range $[0.01,1/6]$.
We used ``fminbnd" to minimize $R(\lambda_1)^2$ as well, but in two steps: first we searched $\lambda_1$ in the range $[0,\pi/2]$ and then in the range $[\pi/2,\pi]$.

For both, MC and $\tau$-leaping, the condition that determines whether the particle is inside or outside of the VOI is as follows:
\begin{eqnarray}
&&\text{If}\,\,|{\bf g}(\theta')|-|{\bf x}|>0,\,\,\,\,\,\text{particle inside}\nonumber\\
&&\text{If}\,\,|{\bf g}(\theta')|-|{\bf x}|\le 0,\,\,\,\,\,\text{particle outside},\nonumber
\end{eqnarray}
where ${\bf x}$ is the particle's position vector and $\theta'$ is its polar angle, which can be computed from its components $(x,y,z)$:
\begin{equation}
  \theta'=\left\{
  \begin{array}{@{}ll@{}}
    \arctan\frac{\sqrt{x^2+y^2}}{z}, & \text{if}\ z>1 \\
    \pi+\arctan\frac{\sqrt{x^2+y^2}}{z}, & \text{if}\ z<0 \\
    \frac{\pi}{2}, & \text{if}\ z=0 \ \text{and} \ x\neq y\neq 0. 
  \end{array}\right.
\end{equation} 

\subsection{Example 2}

Let us now generate a more complicated volume by allowing the length of the vector ${\bf g}$ to depend on $\lambda_2$ as well: 
\begin{equation}
{\bf g}(\lambda_1,\lambda_2)=(h(\lambda_1)f(\lambda_2)\sin\lambda_1\cos\lambda_2,h(\lambda_1)f(\lambda_2)\sin\lambda_1\sin\lambda_2,h(\lambda_1)\cos\lambda_1)
\end{equation}
where
\begin{equation}
f(\lambda_2)=1-\frac{\cos(4\lambda_2)}{4}.
\end{equation}
The corresponding volume is shown in Fig. 4. To find the unit vector ${\bf u}(\lambda_1,\lambda_2)$ perpendicular to the surface, we can vary ${\bf g}(\lambda_1,\lambda_2)$ in an arbitrary direction and
demand that
\begin{figure}
	\centering
	\includegraphics[trim=0 0 0 1.0cm, height=0.27\textheight]{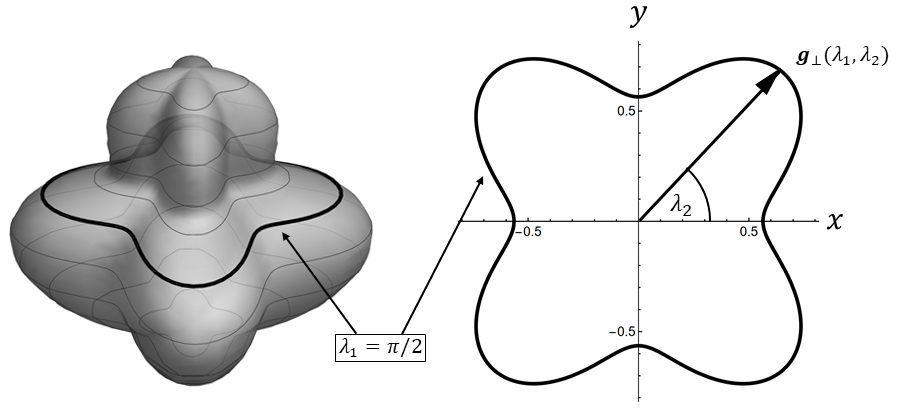}
	\caption{Volume of interest (left) and a horizontal cross-section at $\lambda_1=\pi/2$ (right).}
\end{figure}
\begin{equation}\label{variation_g}
\delta{\bf g}(\lambda_1,\lambda_2)\cdot{\bf u}(\lambda_1,\lambda_2)
=\frac{\partial{\bf g}(\lambda_1,\lambda_2)}{\partial\lambda_1}\cdot{\bf u}(\lambda_1,\lambda_2)\delta\lambda_1+\frac{\partial{\bf g}(\lambda_1,\lambda_2)}{\partial\lambda_2}\cdot{\bf u}(\lambda_1,\lambda_2)\delta\lambda_2=0.
\end{equation}
Since $\delta\lambda_1$ and $\delta\lambda_2$ are arbitrary, albeit infinitesimal, each of the two terms on the right in Eq. (\ref{variation_g}) must be zero. Hence,
\begin{eqnarray}
&&\frac{\partial{\bf g}(\lambda_1,\lambda_2)}{\partial\lambda_1}\cdot{\bf u}(\lambda_1,\lambda_2)=0\nonumber\\
&&\frac{\partial{\bf g}(\lambda_1,\lambda_2)}{\partial\lambda_2}\cdot{\bf u}(\lambda_1,\lambda_2)=0\nonumber,
\end{eqnarray}
which, when coupled with the condition that $u_x^2+u_y^2+u_z^2=1$, yields a unique solution to $u_x$, $u_y$ and $u_z$:
\begin{eqnarray}
&&u_x(\lambda_1,\lambda_2)=[h(\lambda_1)\sin\lambda_1-\cos\lambda_1h'(\lambda_1)] [\cos\lambda_1f(\lambda_2)+\sin\lambda_1f'(\lambda_2)]/K(\lambda_1,\lambda_2)\nonumber\\
&&u_y(\lambda_1,\lambda_2)=[h(\lambda_1)\sin\lambda_1-\cos\lambda_1h'(\lambda_1)] [f(\lambda_2)\sin\lambda_2-\cos\lambda_2f'(\lambda_2)]/K(\lambda_1,\lambda_2)\nonumber\\
&&u_z(\lambda_1,\lambda_2)=[f(\lambda_2)^2 (\cos\lambda_1h(\lambda_1)+\sin\lambda_1h'(\lambda_1)]/K(\lambda_1,\lambda_2),\nonumber
\end{eqnarray}
where
\begin{eqnarray}
&&K(\lambda_1,\lambda_2)=\nonumber\\
&&\left\{f(\lambda_1)^4[\cos\lambda_1 h(\lambda_1)+\sin\lambda_1 h'(\lambda_1)]^2 + [h(\lambda_1)\sin\lambda_1-\cos\lambda_1h'(\lambda_1)]^2[f(\lambda_2)^2+f'(\lambda_2)^2]\right\}^{1/2}.\nonumber\\
\end{eqnarray}
The square radius to be minimized is now
\begin{eqnarray}
R(\lambda_1,\lambda_2)^2&=&[h(\lambda_1)f(\lambda_2)\sin\lambda_1\cos\lambda_2-R_{\text{min}}u_x(\lambda_1,\lambda_2)]^2\nonumber\\
&+&[h(\lambda_1)f(\lambda_2)\sin\lambda_1\sin\lambda_2-R_{\text{min}}u_y(\lambda_1,\lambda_2)]^2\nonumber\\
&+&[h(\lambda_1)\cos\lambda_1-R_{\text{min}}u_y(\lambda_1,\lambda_2)]^2.
\end{eqnarray}

To minimize $R(\lambda_1,\lambda_2)^2$, we formed a grid by dividing $\lambda_1$ into two sections - $[0,\pi/2]$ and $[\pi/2,\pi]$ -  and $\lambda_2$ into five sections -$[0,2\pi/5]$, $[2\pi/5,4\pi/5]$, $[4\pi/5,6\pi/5]$, $[6\pi/5,8\pi/5]$ and
$[8\pi/5,2\pi]$ - and used ``fmincon", with the initial search point being in the middle of each pixel.

The condition that determines whether the particle is inside or outside of the VOI is now a function of two variables:
\begin{eqnarray}
&&\text{If}\,\,|{\bf g}(\theta',\phi')|-|{\bf x}|>0,\,\,\,\,\,\text{particle inside}\nonumber\\
&&\text{If}\,\,|{\bf g}(\theta',\phi')|-|{\bf x}|\le 0,\,\,\,\,\,\text{particle outside},\nonumber
\end{eqnarray}
where ${\bf x}$ is the particle's position vector and $\theta'$ and $\phi'$ are its polar and azimuthal angles, and can be computed from its components $(x,y,z)$:
\begin{eqnarray}
  &&\phi'=\left\{
  \begin{array}{@{}ll@{}}
    \arcsin\frac{y}{\sqrt{x^2+y^2}}, & \text{if}\ x>0 \ \text{and}\ y>0 \\
    \pi-\arcsin\frac{y}{\sqrt{x^2+y^2}}, & \text{if}\ x<0 \ \text{and}\ y\neq 0 \\
    2\pi+\arcsin\frac{y}{\sqrt{x^2+y^2}}, & \text{if}\ x>0 \ \text{and}\ y<0 \\ 
  \end{array}\right.
\\
  &&\theta'=\left\{
  \begin{array}{@{}ll@{}}
    \arctan\frac{\sqrt{x^2+y^2}}{zf(\phi')}, & \text{if}\ z>0 \\
    \pi+\arctan\frac{\sqrt{x^2+y^2}}{zf(\phi')}, & \text{if}\ z<0 \\
    \frac{\pi}{2}, & \text{if}\ z=0 \ \text{and} \ x\neq y\neq 0. 
  \end{array}\right.
\end{eqnarray}

\subsection{Results}

\begin{figure}\label{Results1}
	\centering
	\includegraphics[trim=0 0 0 1.0cm, height=0.42\textheight]{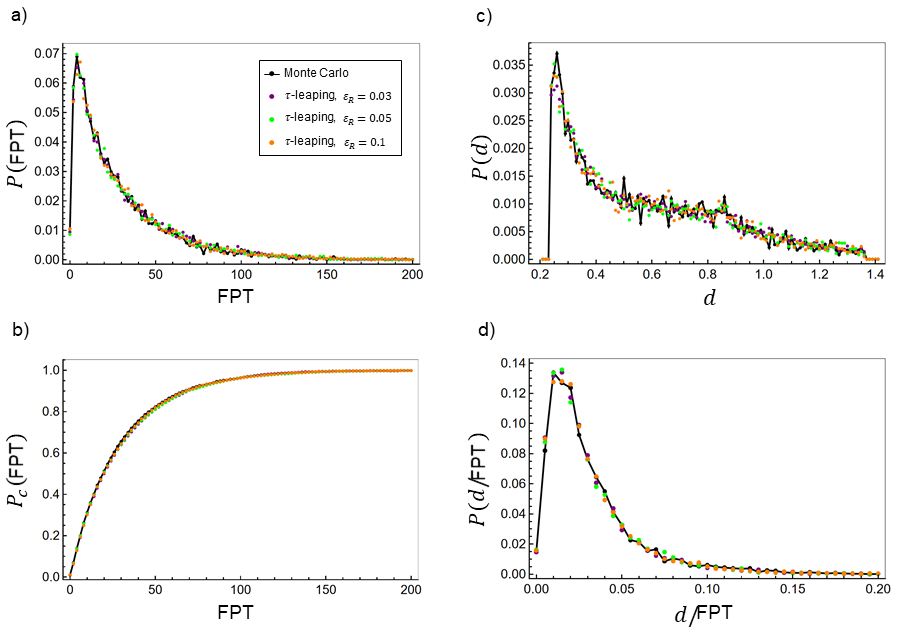}
	\caption{{\bf Example volume 1:} Monte Carlo (black) and $\tau$-leaping for three values of $\varepsilon_R$ - 0.03 (purple), 0.05 (green) and 0.1 (orange) - for a) probability for FPT;
b) cumulative probability for the FPT; c) probability for the distance between the initial position and the point of crossing; and d) probability for the speed, i. e. distance between the initial position and the point of crossing divided by
the FPT. The bin sizes are: a) 1, b) 1, c) 0.005, and d) 0.01.}
\end{figure}

\begin{figure}\label{Results1}
	\centering
	\includegraphics[trim=0 0 0 1.0cm, height=0.42\textheight]{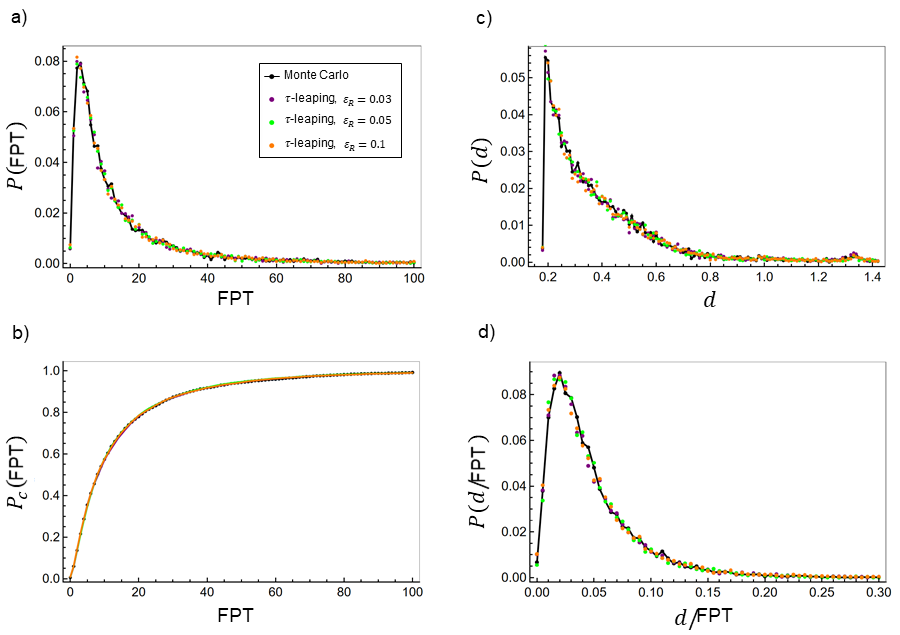}
	\caption{{\bf Example volume 2:} Monte Carlo (black) and $\tau$-leaping for three values of $\varepsilon_R$ - 0.03 (purple), 0.05 (green) and 0.1 (orange) - for a) probability for FPT;
b) cumulative probability for the FPT; c) probability for the distance between the initial position and the point of crossing; and d) probability for the speed, i. e. distance between the initial position and the point of crossing divided by
the FPT. The bin sizes are: a) 1, b) 1, c) 0.005, and d) 0.01.}
\end{figure}

The parameter values for all simulations were chosen
to be: $k_BT=4.14\times 10^{-9}$kg$\cdot\mu$m$^2\cdot$s$^{-2}$, $m=10^{-10}$kg, viscosity $\nu=1.7\times 10^{-9}$kg$\cdot\mu$m$^{-1}\cdot$s$^{-1}$, and particle's size $r_B=58.6\mu$m. 
These values render the relaxation time $\tau_B=5.31\times 10^{-5}$s. In all simulations, $dt$ was chosen to be $5\times 10^{-6}$s. In the two examples above, the initial positions were chosen to be $(0,0.4,0)$ and
$(0.5,0.5,0)$ respectively. 
Figures 5 and 6 shows the comparisons between Monte Carlo and the $\tau$-leaping method for example volumes 1 and 2, respectively. 

\section{Discussion}

We have presented a $\tau$-leaping method to compute the first passage time (FPT) and position of a Brownian particle. The ``leaping" was done by sampling the FPT and position for a sphere inscribed in the volume of interest (VOI) and centered
at the last sampled position of the particle. By setting a lower limit on the size of such a sphere, $R_{\text{min}}$, and repeating the ``leaping" procedure, we eventually arrive at a position 
(near the surface of the VOI) where the size of the sphere is less than $R_{\text{min}}$;
at such a point, the method switches to regular Monte Carlo simulation until the particle either leaves the VOI, or reaches a position where a sphere of radius greater than $R_{\text{min}}$ can be generated. The purpose of setting a lower limit on
the size of the spheres was to avoid having to sample the velocity of the particle: the larger the sphere, the less important the initial velocity for the sampling of FPT and position. Hence, $R_{\text{min}}$ is chosen based on one's notion of accuracy.
Another important step in this method is to give the VOI an inner skin of thickness $R_{\text{min}}$. This, again, is to avoid having to sample velocities: by generating spheres that are inscribed by the volume bounded by the inner
surface of the skin, we are guaranteed (within an accuracy we have chosen by setting $R_{\text{min}}$) that the particle's velocity at the last sampling will not be important in the Monte Carlo simulation when the particle evolves to a distance
greater than or equal to $R_{\text{min}}$. We have demonstrated this method, on two example volumes and three thicknesses of skin to be as accurate and much more efficient than Monte Carlo, as shown in Table 1. The last column
gives the percentage values of the average distance between the probabilities for the FPT of Monte Carlo and $\tau$-leaping:
\begin{equation}
\text{Accuracy}=100\left[1-\sum_{n=1}^{N_t}|P_{\tau}(T_n)-P_{MC}(T_n)|/N_t\right],
\end{equation}
where $N_t$ is the number of bins in the histograms in Figures 5a and 6a.
Although the accuracy for the three choices of $\varepsilon_R$ is essentially the same, the efficiency varies significantly. According to the condition
$t/t_{\text{min}}=6\times0.04(\varepsilon_d/\varepsilon_R^2)\gg 1$ (see the last paragraph of section ``Brownian motion and the Langevin equation"), the three values of 
$\varepsilon_R$, 0.03, 0.05 and 0.1, give $t/t_{\text{min}}=$8, 2.88 and 0.72, respectively, only the first of which can be said to satisfy the condition $t/t_{\text{min}}\gg 1$. What this tells us
is that the condition itself might be too strict and further analysis is needed to refine it.

\begin{table}[htbp]
	\centering
	\caption{Values for efficiency of Monte Carlo simulations and the $\tau$-leaping method (column 4) as a function of volume of interest and $R_{\text{min}}$. Column 5 shows the accuracy of the $\tau$-leaping method relative to
Monte Carlo.}
\begin{tabular}{lclclclcl}
	\hline	
	\multicolumn{1}{|l|}{\multirow{2}{*}{\,\,\,\,\,\,\,\,\,Method}} &\multirow{2}{*}{\,\,Volume \#\,\,} &\multicolumn{1}{|l|}{\multirow{2}{*}{\,\,$\varepsilon_r$,\,$R_{\text{min}}\,(\mu$m)}\,\,}& \,\,\,\,\,Average efficiency\,\,\, & \multicolumn{1}{|l}{\multirow{2}{*}{\,\,Accuracy}} & \multicolumn{1}{l|}{} \\
	\multicolumn{1}{|l|}{}	& & \multicolumn{1}{|l|}{}&  (seconds/run)  & \multicolumn{1}{|l}{} & \multicolumn{1}{l|}{} \\
	\hline
	\hline
	\multicolumn{1}{|l|}{\multirow{2}{*}{\,\,Monte Carlo\,\,}}     & 1 & \multicolumn{1}{|l|}{\,\,\,\,\,\,\,\,\,\,\,\,NA}  & 155.52 & \multicolumn{1}{|l}{\,\,\,\,\,\,\,\,\,NA} & \multicolumn{1}{l|}{} \\  \cline{2-6}
	\multicolumn{1}{|l|}{\,\,\,\,\,\,\,\,\,\,\,\,\,\,\,\,\,\,\,\,\,\,\,\,\,\,\,\,\,\,\,\,\,}                                             & 2 & \multicolumn{1}{|l|}{\,\,\,\,\,\,\,\,\,\,\,\,NA}  & 156.25 &  \multicolumn{1}{|l}{\,\,\,\,\,\,\,\,\,NA} & \multicolumn{1}{l|}{} \\	
	\hline
	\multicolumn{1}{|l|}{\multirow{6}{*}{\,\,\,\,\,\,$\tau$-leaping\,\,\,}}&  & \multicolumn{1}{|l|}{\,\,\,\,\,0.03,\,\,\,0.034} & 12.10 &  \multicolumn{1}{|l}{\,\,\,\,99.917\%} & \multicolumn{1}{l|}{} \\ 
	\multicolumn{1}{|l|}{}		 & 1 & \multicolumn{1}{|l|}{\,\,\,\,\,\,0.05,\,\,\,0.02} & 1.84 &  \multicolumn{1}{|l}{\,\,\,\,99.914\%} & \multicolumn{1}{l|}{} \\   
	\multicolumn{1}{|l|}{}	        &   & \multicolumn{1}{|l|}{\,\,\,\,\,\,\,0.1,\,\,\,0.01}  & 1.38  &  \multicolumn{1}{|l}{\,\,\,\,\,\,99.9\%} & \multicolumn{1}{l|}{} \\ \cline{2-6}  
	\multicolumn{1}{|l|}{}		&    & \multicolumn{1}{|l|}{\,\,\,\,\,0.03,\,\,\,0.034} & 16.9 &  \multicolumn{1}{|l}{\,\,\,\,99.929\%} & \multicolumn{1}{l|}{} \\ 					      
  	\multicolumn{1}{|l|}{}		& 2  & \multicolumn{1}{|l|}{\,\,\,\,\,\,0.05,\,\,\,0.02} & 13.69  &  \multicolumn{1}{|l}{\,\,\,\,99.911\%} & \multicolumn{1}{l|}{} \\  					      
 	\multicolumn{1}{|l|}{}		 &    & \multicolumn{1}{|l|}{\,\,\,\,\,\,\,0.1,\,\,\, 0.01}  & 12.0  &  \multicolumn{1}{|l}{\,\,\,\,99.904\%} & \multicolumn{1}{l|}{} \\           
	\hline
\end{tabular} 
\end{table}

We should point out that the size of the particle we have chosen as our test subject was $\sim60\mu$m, while the enclosing volumes were $\sim 1\mu$m large. This may seem like a geometric impossibility; however, it is not, since the volumes
are imaginary and only serve to facilitate a comparison between two methods. A more realistic scenario would have been to chose a volume much larger than the particle's size, in which case the volume could be treated as a real physical enclosure. However, this would make Monte Carlo simulations infeasible: for a volume 10 times larger than the particle's radius ($\sim600\mu$m) 1000 simulations would take about 4.5$\times 10^34$ hours. On the other hand, because the efficiency of our method is hindered only by the thickness of the skin, which does not change with scaling of the volume, it would
be effected hardly at all. Another realistic scenario would have been to make the Brownian particle much smaller, while keeping the volumes fixed. For example, mass and viscosity typical of biological cells,
$m=10^{-20}$kg, and $\nu=1.7\times 10^{-8}$kg$\cdot\mu$m$^{-1}\cdot$s$^{-1}$, and an average protein size $\sim5.86\times 10^{-4}\mu$m, would give $\tau_B=5.31\times 10^{-11}$s
and the values for $R_{\text{min}}$ ten times smaller than used in this paper, which would make the $\tau$-leaping method faster still by a factor of $\sim$10.


The relatively simple structure of our method makes it ideal for simulations that combine interactions of a particle with not only boundaries, but also objects within the boundaries. For example, a protein, 
seeking a binding site on DNA, would typically bounce or slide along the chromatin, thus effectively reducing the search space from three to two (or even one, for unwound chromatin) dimensions.
Our method can be easily applied in this scenario by simply generating a skin around the chromatin.


\end{document}